# Sentiment Analysis and Text Analysis of the Public Discourse on Twitter about COVID-19 and MPox

**Nirmalya Thakur**

Department of Computer Science, Emory University, Atlanta, GA 30322, USA; nirmalya.thakur@emory.edu

**Abstract:** Mining and analysis of the big data of Twitter conversations have been of significant interest to the scientific community in the fields of healthcare, epidemiology, big data, data science, computer science, and their related areas, as can be seen from several works in the last few years that focused on sentiment analysis and other forms of text analysis of tweets related to Ebola, E-Coli, Dengue, Human Papillomavirus (HPV), Middle East Respiratory Syndrome (MERS), Measles, Zika virus, H1N1, influenza-like illness, swine flu, flu, Cholera, Listeriosis, cancer, Liver Disease, Inflammatory Bowel Disease, kidney disease, lupus, Parkinson's, Diphtheria, and West Nile virus. The recent outbreaks of COVID-19 and MPox have served as "catalysts" for Twitter usage related to seeking and sharing information, views, opinions, and sentiments involving both of these viruses. None of the prior works in this field analyzed tweets focusing on both COVID-19 and MPox simultaneously. To address this research gap, a total of 61,862 tweets that focused on MPox and COVID-19 simultaneously, posted between 7 May 2022 and 3 March 2023, were studied. The findings and contributions of this study are manifold. First, the results of sentiment analysis using the VADER (Valence Aware Dictionary for sEntiment Reasoning) approach shows that nearly half the tweets (46.88%) had a negative sentiment. It was followed by tweets that had a positive sentiment (31.97%) and tweets that had a neutral sentiment (21.14%), respectively. Second, this paper presents the top 50 hashtags used in these tweets. Third, it presents the top 100 most frequently used words in these tweets after performing tokenization, removal of stopwords, and word frequency analysis. The findings indicate that tweets in this context included a high level of interest regarding COVID-19, MPox and other viruses, President Biden, and Ukraine. Finally, a comprehensive comparative study that compares the contributions of this paper with 49 prior works in this field is presented to further uphold the relevance and novelty of this work.

## 1. Introduction

In today's world, social media serves as an "integral vehicle" [1] and as an "online community" [2] for seeking and sharing information, news, views, opinions, perspectives, ideas, awareness, comments, and experiences on various topics, such as pandemics, global affairs, current technologies, recent events, politics, family, relationships, and career opportunities, just to name a few [3]. Out of multiple social media platforms, Twitter is highly popular amongst all age groups. As of December 2022, Twitter's audience accounted for over 368 million monthly active users worldwide [4]. Twitter is the most used social media platform amongst journalists [5] and ranks amongst the most popular social media platforms on a global scale [6]. Twitter has been highly popular amongst data scientists and computer science researchers for studying, analyzing, modeling, and interpreting social media communications related to various topics, such as ChatGPT [7], the Russia–Ukraine war [8], cryptocurrency markets [9], virtual assistants [10], mental health [11], loneliness in the elderly [12], housing needs of low-income families [13], animal welfare [14], climate change [15], cognitive impairment [16], the electronics industry [17], agriculture [18], race and ethnicity [19], fake news [20], abortion [21], religion [22], fall detection [23,24], gender identity [25], elections [26], politics [27], food insufficiency [28], pregnancy [29], drug safety [30], indoor localization [31], gambling [32], education systems [33], exoskeletons [34], personalized medicine [35], natural disasters [36], crimes [37], democracy [38], and transportation [39], just to name a few. In addition to the above, Twitter data mining and analysis has also attracted the attention of healthcare researchers, epidemiologists, and medical practitioners, as is evident from several works that focused on the mining and analysis of tweets related to pandemics, epidemics, viruses, and diseases such as Ebola [40], E-Coli [41], Dengue [42], Human Papillomavirus (HPV) [43], Middle East Respiratory Syndrome (MERS) [44], Measles [45], Zika virus [46], H1N1 [47], influenza-like illness [48], swine flu [49], flu [50], Cholera [51], Listeriosis [52], cancer [53], Liver Disease [54], Inflammatory Bowel Disease [55], kidney disease [56], lupus [57], Parkinson's [58], Diphtheria [59], and West Nile virus [60].

The recent outbreaks of COVID-19 and MPox have served as "catalysts", leading to the usage of Twitter for the sharing and exchanging information on diverse topics related to these viruses, leading to the generation of tremendous amounts of big data. No prior work in this field has focused on studying and analyzing tweets that focused on both of these viruses simultaneously to understand and interpret the underlying paradigms of conversations. Therefore, this serves as the main motivation for this work.

In December 2019, there was an outbreak of an unknown respiratory disease in a seafood market in Wuhan, China. This outbreak affected about 66% of the people in the market. A prompt investigation from the healthcare and medical sectors revealed that a novel coronavirus was responsible for this disease, and this virus was named severe acute respiratory syndrome coronavirus-2 (SARS-CoV-2, 2019-nCoV), as it was observed to have a high homology of about 80% with SARS-CoV [61]. The disease that humans suffer from after getting infected by this virus is known as COVID-19 [62]. Despite the best efforts of the Chinese Government to contain the spread of this virus, it soon spread to other parts of the world while undergoing multiple mutations, and several variants, such as Alpha (B.1.1.7), Beta (B.1.351), Gamma (P.1), Delta (B.1.617.2), Epsilon (B.1.427 and B.1.429), Eta (B.1.525), Iota (B.1.526), Kappa (B.1.617.1), Zeta (P.2), Mu (B.1.621 and B.1.621.1), and Omicron (B.1.1.529, BA.1, BA.1.1, BA.2, BA.3, BA.4, and BA.5) [63] led to an increase in COVID-19 cases. At present, there have been a total of 681,518,412 cases and 6,811,869 deaths on account of COVID-19 on a global scale [64]. Respiratory systems are the primary target of the SARS-CoV-2 virus, although infections in other organs of the body have been reported in some cases. The symptoms of COVID-19 usually include fever, dry cough, dyspnea, headache, dizziness, exhaustion, vomiting, and diarrhea [65]. However, studies have shown that symptoms can vary from person to person based on user diversity, such as age group, preexisting conditions, disabilities, etc. [66,67].

MPox (monkeypox) is a re-emerging zoonotic disease. It is caused by the MPox (monkeypox) virus, which belongs to the Poxviridae family, the Chordopoxvirinae subfamily, and the Orthopoxvirus genus [68]. This virus was originally identified in monkeys in 1958 [69], and the first case of this virus in humans was recorded in 1970. The MPox virus is closely related to the variola virus and causes a smallpox-like disease in humans. The common symptoms of MPox include fever, headache, and myalgia. A distinguishing feature of MPox is the presence of swelling at the maxillary, cervical, or inguinal lymph nodes [70,71]. The MPox virus was endemic in the Democratic Republic of the Congo (DRC) and a few African countries for a long time, and a few cases outside these geographic regions were recorded only twice—first in 2003 [72] and then in 2017–2018 [73,74]. However, since May 2022, the world has been experiencing an outbreak of the MPox virus. At present, there have been a total of 86,231 cases of MPox, with 84,858 of these cases being recorded in regions that have not historically reported MPox [75].

In the context of recent works related to Twitter data mining and analysis, a number of works have focused on the sentiment analysis of tweets. Sentiment analysis [76] is the computational analysis of people's attitudes, views, and sentiments regarding an entity, which may represent an individual, concept, topic, event, or scenario. Sentiment analysis can be considered a classification process. The three primary classification levels in sentiment analysis are the document level, sentence level, and aspect level. The goal of document-level sentiment analysis is to categorize an opinion document as expressing a positive or negative sentiment. The entire document is viewed as a single fundamental informational unit in this process. Sentence-level sentiment analysis seeks to categorize the sentiment that each sentence expresses. In order to categorize the sentiment in relation to particular features of entities, aspect-level sentiment analysis is used. While there have been prior works in this field, however, those works focused on performing sentiment analysis of tweets about COVID-19 or MPox and did not perform sentiment analysis of tweets that focused on both of these viruses simultaneously. The outbreak of MPox during the ongoing outbreak of COVID-19 has resulted in several tweets involving the views, opinions, concerns, and perspectives of the public regarding both of these viruses. A few examples of such tweets (obtained by using the Advanced Search feature of Twitter) are shown in Table 1. As can be seen from these tweets, these two ongoing virus outbreaks prompted the sharing and exchange of views, information, concerns, and perspectives on a wide range of topics (related and unrelated to these viruses) that reflect sentiments of varying polarities related to those topics. No prior work in this field thus far has focused on studying and analyzing tweets that involved conversations about both COVID-19 and MPox. This work aims to address this research gap in this field. The work in this paper involved performing sentiment analysis and text analysis on 61,862 tweets that focused on MPox and COVID-19 simultaneously, posted between 7 May 2022 and 3 March 2023. The findings and contributions of this paper are summarized as follows:

- The results of sentiment analysis using the VADER (Valence Aware Dictionary for sEntiment Reasoning) approach shows that nearly half the tweets (46.88%) had a negative sentiment. It was followed by tweets that had a positive sentiment (31.97%) and tweets that had a neutral sentiment (21.14%), respectively.
- Using concepts of text analysis, the top 50 hashtags associated with these tweets were obtained. These hashtags are presented in this paper.
- The top 100 most frequently used words that featured in these tweets were obtained after performing tokenization, removal of stopwords, and word frequency analysis of these tweets. The findings show that some of the commonly used words involved Twitter users directly referring to either or both of these viruses. In addition to this, the presence of words such as "Polio", "Biden", "Ukraine", "HIV", "climate", and "Ebola" in the list of the top 100 most frequent words indicates that topics of conversations on Twitter in the context of COVID-19 and MPox also included a high level of interest related to other viruses, President Biden, and Ukraine.

**Table 1.** A random collection of 10 Tweets that focused on COVID-19 and MPox simultaneously.

| | **Tweets related to COVID-19 and MPox** |
|---|---|
| Tweet #1 | They cant figure out how Monkey Pox got here without traveling, and why people are susceptible to it after Covid? Try looking at your immune system after taking the vaccines. Every disease that ever was, is now something for you to fear. Your immune system has been compromised. |
| Tweet #2 | Thanks all you biden fans letting in all these illegal immigrants that have been coming every day since Biden took office. Now we have to worry even more about a new virus coming into this country Monkey Pox forget Covid welcome MONKEY POX |
| Tweet #3 | So, I've got my rainbow sticker, Thank you NHS on my window,' I've had my covid vaccine' on my fb page, Ukraine flag in the garden. It still isn't enough to show how nice I am! Just need a monkey pox sticker. Deffo going to heaven. Stay safe everyone |
| Tweet #4 | MONKEY POX, I am so not ready for you to show up anywhere. |
| | Can you imagine the dilemma of future docs, Now with long COVID , long monkey, monkey heart, monkey lungs, monkey brain might emerge. a monkey mask might help. If lived long enough, might have COVID docs, monkey docs etc |
| Tweet #5 | Are you kidding me, now Monkey Pox?! I've spent 3 years caring for my ill wife, fighting against Covid, and trying to survive...now this?! Some days... |
| Tweet #6 | I sure hope the Government doesn't plan to try to force everyone to get monkey pox vaccines. I'd hate to see where that goes so shortly after covid. |
| Tweet #7 | Another lockdown is incoming. They are trying to make monkey pox look like a pandemic. Their media tools are ready, their vaccines were ready before the pox was introduced. These were the same people that played the COVID19 play. They just changed the name of the movie. Failure! |
| Tweet #8 | Monkey Pox new Covid. Election is coming. Coincidence? No |
| Tweet #9 | First it was maga. Then there came covid. Now, it's Monkey Pox. When will these horrors end?!? |
| Tweet #10 | No longer scared of disease be it Covid or Monkey pox; I'm scared of loosing more years of my life ... |

[1] These Tweets are presented here in "as is" form after obtaining the same from the Advanced Search feature of Twitter. These Tweets do not represent or reflect the views or opinions or beliefs or political stance of the author of this paper.

In addition to the above, a comprehensive comparative study that compares the contributions of this paper with 49 prior works in this field to uphold its relevance and novelty is also presented in this paper. This paper is organized as follows. In Section 2, an overview of recent works related to sentiment analysis and text analysis of tweets about COVID-19 and MPox is presented. Section 3 discusses the detailed methodology and the specific steps that were followed in this work. In Section 4, the results are presented. Section 5 concludes the paper and outlines the scope for future work in this field. It is followed by references.

## 2. Literature Review

This section is divided into two parts. Section 2.1 presents a review of recent works in this field that focused on sentiment analysis and text analysis of tweets about COVID-19. Section 2.2 presents a review of the recent works in this field that focused on sentiment analysis and text analysis of tweets about MPox.

### *2.1. Recent Works that Focused on Sentiment Analysis and Text Analysis of Tweets about COVID-19*

The study by Vijay et al. [77] examined the impact of COVID-19-related tweets from November 2019 to May 2020 in India. Three categories were created for all tweets (positive, negative, and neutral). To assess how people would react to the COVID-19 lockdown in June 2020, the authors also created many datasets, which were organized by state and month and pooled across all states. The findings showed that most individuals started off tweeting negatively, but as time went on, more and more people began to post positively and neutrally. The work by Mansoor et al. [78] examined the global sentiment analysis of tweets about COVID-19 and the evolution of global sentiment over time. The authors also studied tweets focusing on Work From Home (WFH) and Online Learning to gauge the effect of COVID-19 on daily areas of life. They used different machine-learning models, such as Long Short-Term Memory (LSTM) and Artificial Neural Networks (ANNs), to perform sentiment analysis and reported the accuracy of these models. Pokharel [79] used Google Collab to perform text mining and sentiment analysis of tweets focusing on COVID-19. The study involved collecting tweets posted between 21 May 2020 and 31 May 2020 by Twitter users who shared Nepal as their location. According to the study's findings, while the majority of people had a positive attitude toward COVID-19, there were also situations where fear, grief, and disdain were expressed in the tweets.

In the study by Chakraborty et al. [80], two categories of tweets related to COVID-19 were studied. In one instance, the top 23,000 retweeted tweets over the period of 1 January 2020 to 23 March 2020 were studied. According to the findings presented by the authors, the majority of the tweets expressed neutral or negative emotions. The paper also reports the findings from the analysis of a dataset encompassing 226,668 tweets from the period between December 2019 and May 2020. The findings show that there was a disproportionately high number of neutral and positive tweets posted by internet users. The study also showed that despite the majority of COVID-19-related tweets being positive, internet users were preoccupied with retweeting the negative ones. The objective of the work by Shofiya et al. [81] was to comprehend and examine perceptions of social distancing in the context of COVID-19, as expressed on Twitter. The study focused on analyzing tweets emerging from Canada and containing social-distancing-related keywords. The authors used the SentiStrength tool to determine the sentiment polarity of tweets. Basiri et al. [82] proposed a methodology based on the fusion of four deep-learning models and one classical supervised machine-learning model for sentiment analysis of COVID-19 tweets. They applied their methodology to tweets originating from eight countries. Cheeti et al. [83] used a Naïve Bayes classifier to perform sentiment analysis of tweets focusing on COVID-19, with a specific focus on tweets related to education and learning. In their study, Ridhwan et al. [84] performed sentiment analysis of tweets about COVID-19 posted between 1 February 2020 and 31 August 2020, with a specific focus on tweets that originated from Singapore. The findings showed that the majority of the tweets had positive emotions. Tripathi [85] and Situala et al. [86] used multiple machine-learning approaches to perform sentiment analysis of COVID-19-focused tweets that were posted by people who stated their location as Nepal on Twitter. The purpose of the work by Gupta et al. [87] was to examine the perceptions of Indians, as expressed on Twitter, towards the Indian Government's countrywide lockdown, which was implemented to slow the spread of COVID-19. In this context, the authors used the LinearSVC classifier to perform sentiment analysis, and their classifier achieved a performance accuracy of 84.4%. The work by Alanezi et al. [88] focused on performing sentiment analysis of tweets originating from multiple countries. The results of the study showed that most tweets originating from the USA, Australia, Nigeria, Canada, and England had a neutral sentiment. A similar study that focused on performing sentiment analysis of tweets originating from multiple countries was performed by Dubey [89]. In addition to the above, several studies focused on performing sentiment analyses of tweets about COVID-19 originating from different countries, such as the United Kingdom [90–95], the United States [92,93,96–100], Canada [101–105], India [106–110], Australia [111–113], and Brazil [114–116].

### *2.2. Recent Works that Focused on Sentiment Analysis and Text Analysis of Tweets about MPox*

Iparraguirre-Villanueva et al. [117] aimed to examine people's emotions, including positive, negative, and neutral sentiments, towards the MPox outbreak by analyzing tweets containing the hashtag #Monkeypox. The findings of the study indicated that 45.42% of individuals did not express any discernible positive or negative opinions, whereas 19.45% conveyed negative and apprehensive sentiments related to the outbreak. The objective of the study by Mohbey et al. [118] was to infer the range of reactions of the general public in response to the MPox outbreak. The methodology was based on using CNN and LSTM to study relevant tweets to infer these specific characteristics. Farahat et al. [119] conducted a study involving sentiment analysis and topic modeling of tweets associated with MPox. The tweets that were analyzed in this study were posted on Twitter between 22 May 2022 and 5 August 2022. The authors utilized the concept of keyword search to mine tweets containing the keywords "monkeypox", "Monkeypox cases", and "Monkeypox virus". The findings of the sentiment analysis indicated that 48% of the tweets were neutral, 37% were positive, and 15% were negative. The authors used LDA to extract 12 topics that were present in these tweets. Sv et al. [120] focused on understanding the attitude of the general public towards MPox, as expressed on Twitter. They performed sentiment analysis of tweets containing the keyword "monkeypox" that were posted between 1 June 2022 and 25 June 2022. The results of sentiment analysis showed that the percentage of positive tweets was higher as compared to the percentage of negative tweets. The results of topic detection revealed multiple subject matters associated with both positive and negative tweets. The work by Bengesi et al. [121] was performed primarily in two steps. The first step of their work involved mining relevant tweets related to MPox. Thereafter, in the next step, they developed and used multiple categorization models to perform sentiment analysis of these tweets. Dsouza et al.'s work [122] focused on performing sentiment analysis of specific tweets related to MPox to detect any stigmatization of the LGBTQ+ community on Twitter. They retrieved tweets posted between 1 May 2022 and 7 September 2022 containing the hashtags "#monkeypox", "#MPVS", "stigma", and "#LGBTQ+". The study involved the analysis of a total of 70,832 tweets.

Zuhanda et al. [123] performed sentiment analysis on 5000 tweets about MPox posted on 5 August 2022. The study showed that the terms "health", "emergency", "public", "covid", and "declares" were often used by Twitter users in the context of tweeting about MPox. The NRC lexicon comparison categorization revealed that fear was the most often expressed emotion, with a representation rate of 19.73%. This was followed by sorrow at 14.77%, trust at 13.90%, anger at 9.99%, shock at 9.14%, disgust at 8.12%, and happiness at 7.90%. In the work by Cooper et al. [124], tweets about MPox posted between 1 May 2022 and 23 July 2022 were studied. The results showed that LGBTQ+ advocates or allies posted a total of 48,330 tweets, and the average sentiment score for all the tweets was −0.413 on a scale of −4 to +4. Ng et al. [125] collected tweets that contained "monkeypox" or "monkey pox" posted on Twitter between 6 May 2022 and 23 July 2022. They used concepts of topic modeling and sentiment analysis to infer the characteristics of the communication expressed in these tweets. The authors identified five topics, which they divided into three main themes. These included worries about safety, the stigmatization of minority populations, and a general loss of confidence in governmental institutions. The public sentiments highlighted increasing and existing partisanship, personal health concerns related to the changing situation, and worries about how the media portrayed minority and LGBTQ communities.

As can be seen from these works that focused on sentiment analysis and text analysis of tweets related to MPox and COVID-19, none of them focused on mining and analyzing tweets that focused on COVID-19 and MPox at the same time to infer the underlying patterns of sentiments. The work presented in this paper aims to address this research gap. The methodology that was followed is discussed in Section 3, and the results are presented in Section 4.

## 3. Methodology

This section outlines the methodology that was followed for the development and implementation of the proposed framework for performing sentiment analysis and text analysis of tweets that focused on COVID-19 and MPox simultaneously.

First of all, a relevant Twitter dataset had to be selected. The dataset that was selected for this study is MonkeyPox2022Tweets [126]. This dataset presents more than 600,000 Tweet IDs of tweets about the 2022 outbreak of MPox. These tweets were posted between 7 May 2022 and 3 March 2023. The dataset comprises tweets in 34 languages, with English being the most common language in which the Tweets are available. The tweets in the dataset include 5470 distinct hashtags related to MPox, out of which #monkeypox is the most frequent hashtag. As this dataset comprises only Tweet IDs, the Hydrator app [127] was used to hydrate this dataset. The process of hydration refers to the process of obtaining the tweets and related information corresponding to each of the Tweet IDs. The Hydrator app works by complying with the policies of accessing the Twitter API, as well as the specific rate limits in terms of accessing the Twitter API. The following steps were followed for hydrating the Tweet IDs present in this dataset:

1. The desktop version of Hydrator was downloaded and installed on a computer with a Microsoft Windows 10 Pro operating system (Version 10.0.19043 Build 19043) comprising Intel(R) Core (TM) i7-7600U CPU @ 2.80 GHz, 2904 Mhz, 2 Core(s) and 4 Logical Processor(s)
2. The Hydrator app was then connected to the Twitter API by clicking on the "Link Twitter Account" button on the app's interface.
3. This next step involved uploading a dataset file to the Hydrator app for hydration. As the Hydrator app allows only one file to be uploaded at a time, all the dataset files (containing only Tweet IDs) were merged to create one .txt file, which was uploaded to the app.
4. Then, specific information about the uploaded dataset file (such as Title, Creator, Publisher, and URL) was entered in the Hydrator app, and then the "Add Dataset" button was clicked to complete the process of dataset upload.
5. Thereafter, in the "Datasets" tab of the Hydrator app, the "Start" button was clicked to initiate the process of hydration.

Figure 1 is a screenshot from the Hydrator app obtained after the completion of this hydration task. The output of the Hydrator app provided 509,248 tweets about MPox. Upon obtaining these tweets, it was crucial to perform text filtering to obtain tweets that contained keywords related to COVID-19. The specific keywords that were selected for text filtering were "COVID", "COVID19", "coronavirus", "coronavirus pandemic", "COVID-19", "corona", "corona outbreak", "omicron variant", "SARS-CoV-2", "corona virus", and "Omicron". These keywords were selected based on the findings of [128]. The text filtering task produced a set of 61,862 Tweets, i.e., each of these Tweets focused on MPox and COVID-19 at the same time. This set of 61,862 tweets was selected for performing sentiment analysis and text analysis. It is worth mentioning here that Twitter introduced multiple changes to the Twitter API in April 2023, as a result of which the Hydrator app is not functional at present. However, the work that involved the usage of the Hydrator app was completed by the first week of March 2023. So, the recent changes to the Twitter API did not have any effect on this study.

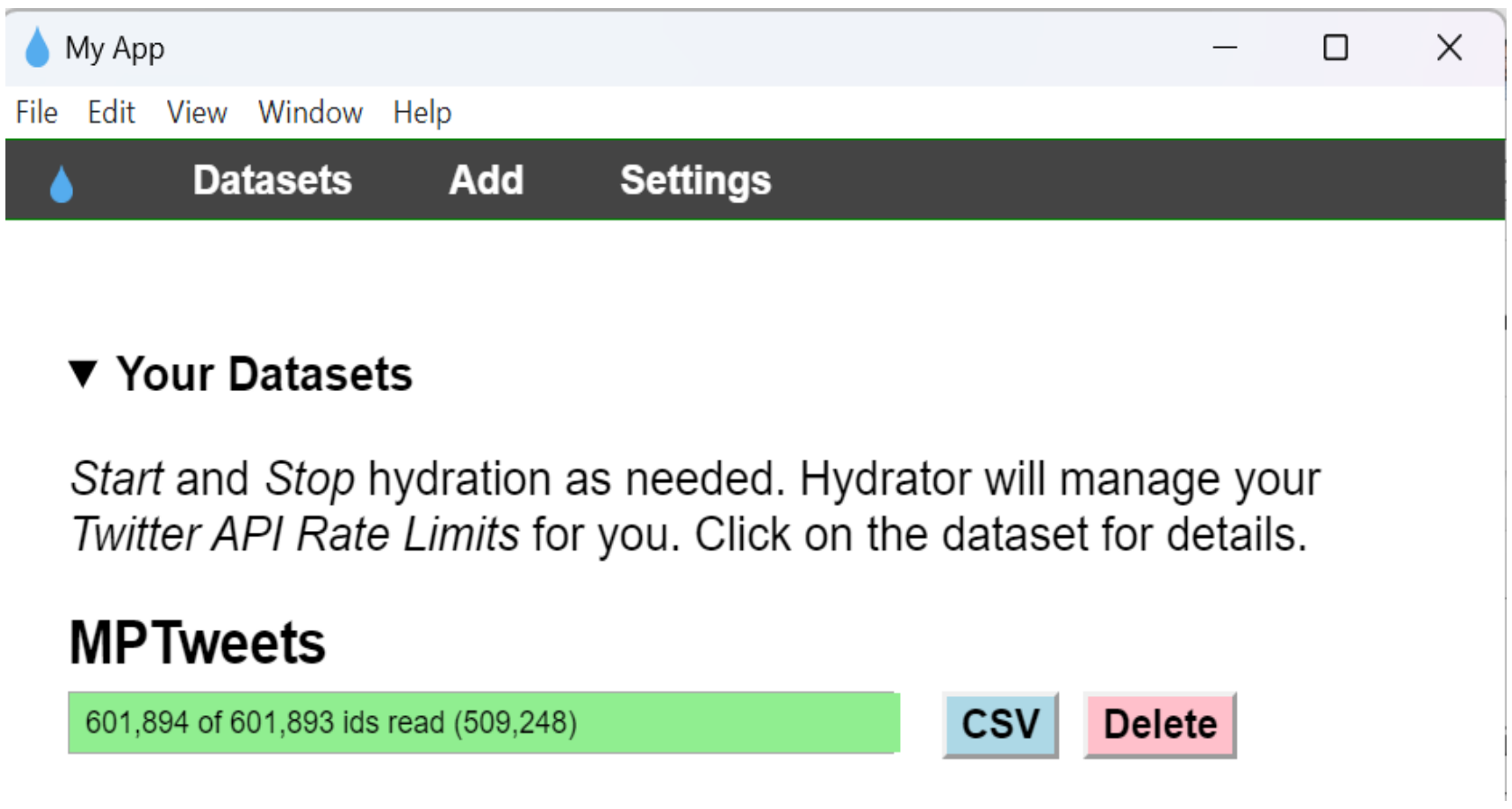

**Figure 1.** Screenshot from the Hydrator app after completion of the Hydration process.

There are various approaches for performing sentiment analysis, such as manual classification, Linguistic Inquiry and Word Count (LIWC), Affective Norms for English Words (ANEW), the General Inquirer (GI), SentiWordNet, and machine-learning-oriented techniques relying on Naïve Bayes, Maximum Entropy, and Support Vector Machine (SVM) algorithms. However, the specific approach that was used in this study was VADER (Valence Aware Dictionary for sEntiment Reasoning). VADER was used because it has been reported to outperform manual classification, and it addresses the limitations in similar approaches for sentiment analysis, as outlined below [129]:

a. VADER distinguishes itself from LIWC, as it is more sensitive to sentiment expressions in social media contexts.
b. The General Inquirer suffers from a lack of coverage of sentiment-relevant lexical features common to social text.
c. The ANEW lexicon is also insensitive to common sentiment-relevant lexical features in social text.
d. The SentiWordNet lexicon is very noisy; a large majority of synsets have no positive or negative polarity.
e. The Naïve Bayes classifier involves the naïve assumption that feature probabilities are independent of one another.
f. The Maximum Entropy approach makes no conditional independence assumption between features and thereby accounts for information entropy (feature weightings).
g. In general, machine-learning classifiers require (often extensive) training data, which are, as with validated sentiment lexicons, sometimes troublesome to acquire.
h. In general, machine-learning classifiers also depend on the training set to represent as many features as possible.

VADER uses sparse-rule-based modeling to build a computational sentiment analysis engine that performs well on the social-media-style text while easily generalizing to multiple domains, needs no training data but is built from a generalizable, valence-based, human-curated gold-standard sentiment lexicon, is quick enough to utilize online with streaming data, does not suffer significantly from a speed–performance tradeoff, has a time complexity of O(N), and is freely available without any subscription or purchase costs. In addition to detecting the polarity (positive, negative, and neutral), VADER is also able to detect the intensity of the sentiment expressed in the texts. To develop the system architecture for sentiment analysis and text analysis, RapidMiner was used. RapidMiner, formerly known as Yet Another Learning Environment (YALE) [130], is a data science platform that enables the development, implementation, and utilization of several algorithms and models related to machine learning, data science, artificial intelligence, and big data. RapidMiner is utilized for both academic research and the creation of business-related applications and solutions. RapidMiner is available as an integrated development environment that consists of (1) RapidMiner Studio, (2) RapidMiner Auto Model, (3) RapidMiner Turbo Prep, (4) RapidMiner Go, (5) RapidMiner Server, and (6) RapidMiner Radoop. For all the work related to the methodologies proposed in this paper, RapidMiner Studio was used. For the remainder of this paper, wherever the phrase "RapidMiner" is used, it refers to "RapidMiner Studio" and not any of the other development environments associated with this software tool. RapidMiner is created as an open-core model with a powerful Graphical User Interface (GUI) that enables developers to create numerous applications and workflows and develop and implement algorithms. In the RapidMiner development environment, specific operations or functions are referred to as "operators" and a collection of "operators" (connected linearly or hierarchically or as a combination of both) to achieve a desired task or goal is referred to as a "process". For the creation of a particular "process", RapidMiner offers a variety of built-in "operators" that may be utilized directly. A particular class of "operators" can also be utilized to change the distinguishing qualities of other "operators". Moreover, the development environment also allows developers to develop their own "operators", which can then be shared and made accessible to all other RapidMiner users via the RapidMiner Marketplace.

The VADER approach for performing sentiment analysis is available as an "operator" in RapidMiner, which can be directly used in a "process". This "operator" calculates and then outputs the sum of all sentiment word scores in a given text(s) by following the VADER approach. If the advanced output option of this "operator" is selected, then it also outputs a nominal attribute with all words taking part in the scoring, the sum of positive components, the sum of negative components, and the number of used and unused tokens. The "process" that was developed in RapidMiner involving the use of this "operator" and other "operators" connected to it is shown in Figure 2.

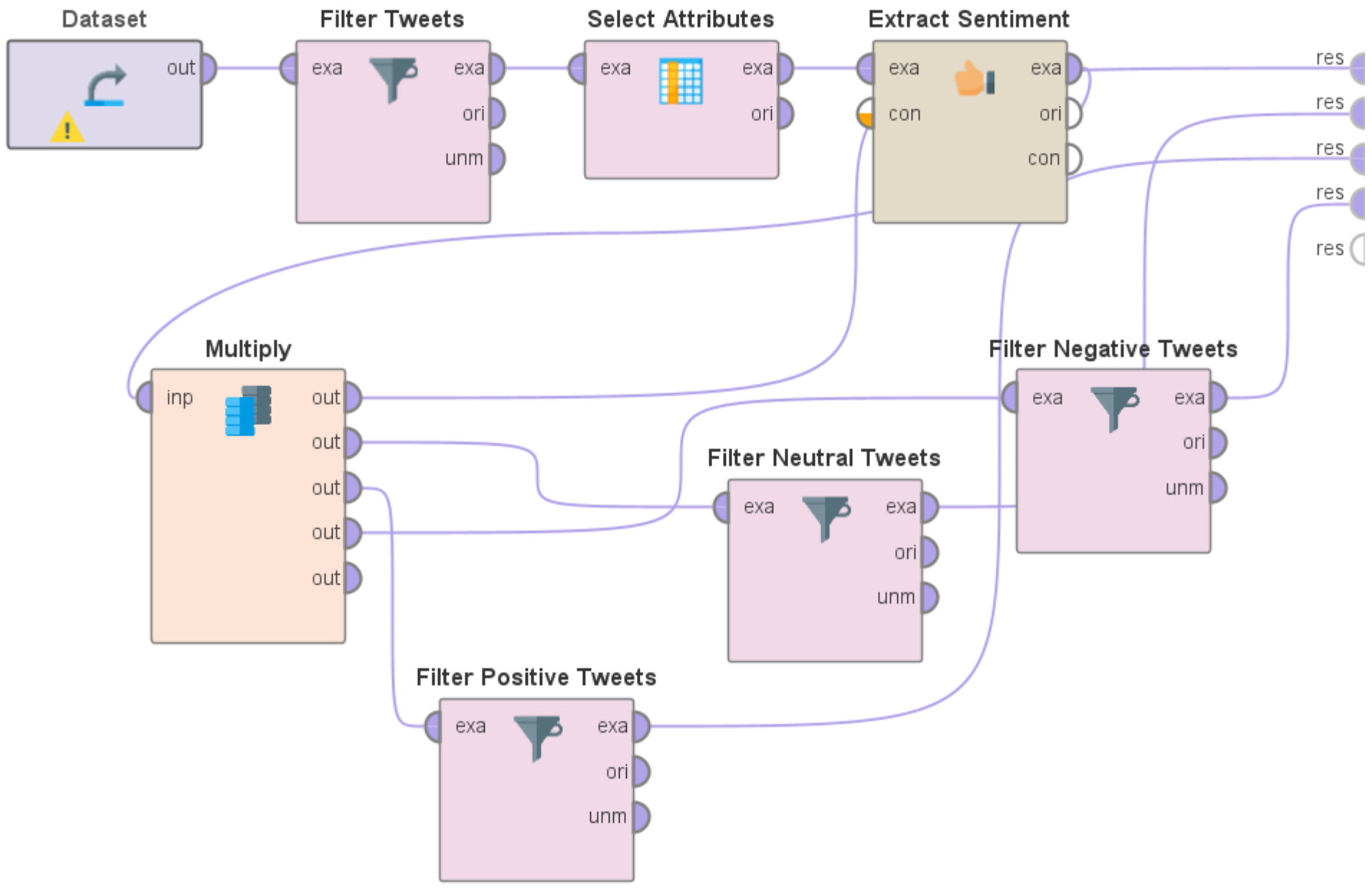

**Figure 2.** The RapidMiner process developed for performing sentiment analysis.

The description of all the "operators" used in this "process" is presented next. The "Dataset" "operator" was used to import the original dataset of 509,248 tweets about MPox (obtained from the output of the Hydrator app). The "Filter Tweets" "operator" was used to perform text filtering on the text of the tweets. Specifically, tweets that contained one or more of these keywords — "COVID", "COVID19", "coronavirus", "coronavirus pandemic", "COVID-19", "corona", "corona outbreak", "omicron variant", "SARS-CoV-2", "corona virus", and "Omicron" were filtered. Thereafter, the "Select Attributes" "operator" was used to select only that specific attribute from the dataset that would be used for sentiment analysis. The specific attribute in this context was the text of the tweets. The output of this "operator" was provided as an input to the "Extract Sentiment" "operator", which performed sentiment analysis according to the VADER approach. The output of this "operator" comprised a score associated with each tweet, classifying it as a positive, neutral, or negative tweet. To compute the number of positive, neutral, or negative tweets, additional data filters were used. However, this required creating multiple copies of the output. To achieve this, the "Multiply" "operator" was used. Specifically, three copies of the output from the VADER "operator" were created by using this operator. These copies of the output were passed through data filters that had been set up to filter out the positive, neutral, and negative tweets based on specific rules as per the working of the VADER approach. These rules were—a tweet with a score greater than 0 was filtered as a positive tweet, a tweet with a score equal to 0 was filtered as a neutral tweet, and a tweet with a score less than 0 was filtered as a negative tweet. Thereafter, an analysis of the number of tweets from these respective data filters was performed to infer the percentages of positive, neutral, and negative tweets. These results are discussed in Section 4.

In addition to performing sentiment analysis, this study also involved the detection of some of the commonly used hashtags and words in the 61,862 tweets that were considered for this study. The RapidMiner "process" that was developed to implement the same is shown in Figure 3. The description of all the "operators" used in this "process" is presented next. The "Dataset" "operator" was used to import the original dataset of 509,248 tweets about MPox (obtained from the output of the Hydrator app). The "Filter Tweets" "operator" was used to perform text filtering on the text of the tweets. Specifically, tweets that contained one or more of these keywords — "COVID", "COVID19", "coronavirus", "coronavirus pandemic", "COVID-19", "corona", "corona outbreak", "omicron variant", "SARS-CoV-2", "corona virus", and "Omicron" were filtered. Thereafter, the "Select Attributes" "operator" was used to select only that specific attribute from the dataset that would be used for sentiment analysis. The specific attribute in this context was the text of the tweets. The output of this "operator" was provided as an input to the "Nominal to Text" operator. Thereafter, the "sub-process" "Process Documents" was used. This "sub-process" comprised specific operators to perform tokenization and elimination of stopwords. The output of this "operator" was provided as an input to the "WordList to Data" operator to display the results for detection and analysis of the commonly used hashtags and words in these tweets. The results of this "process" are also discussed in Section 4. It is worth mentioning here that the VADER "operator" performs tokenization and elimination of stopwords automatically, so the "sub-process" "Process Documents" was not used in the RapidMiner "process" (shown in Figure 2) to perform the sentiment analysis.

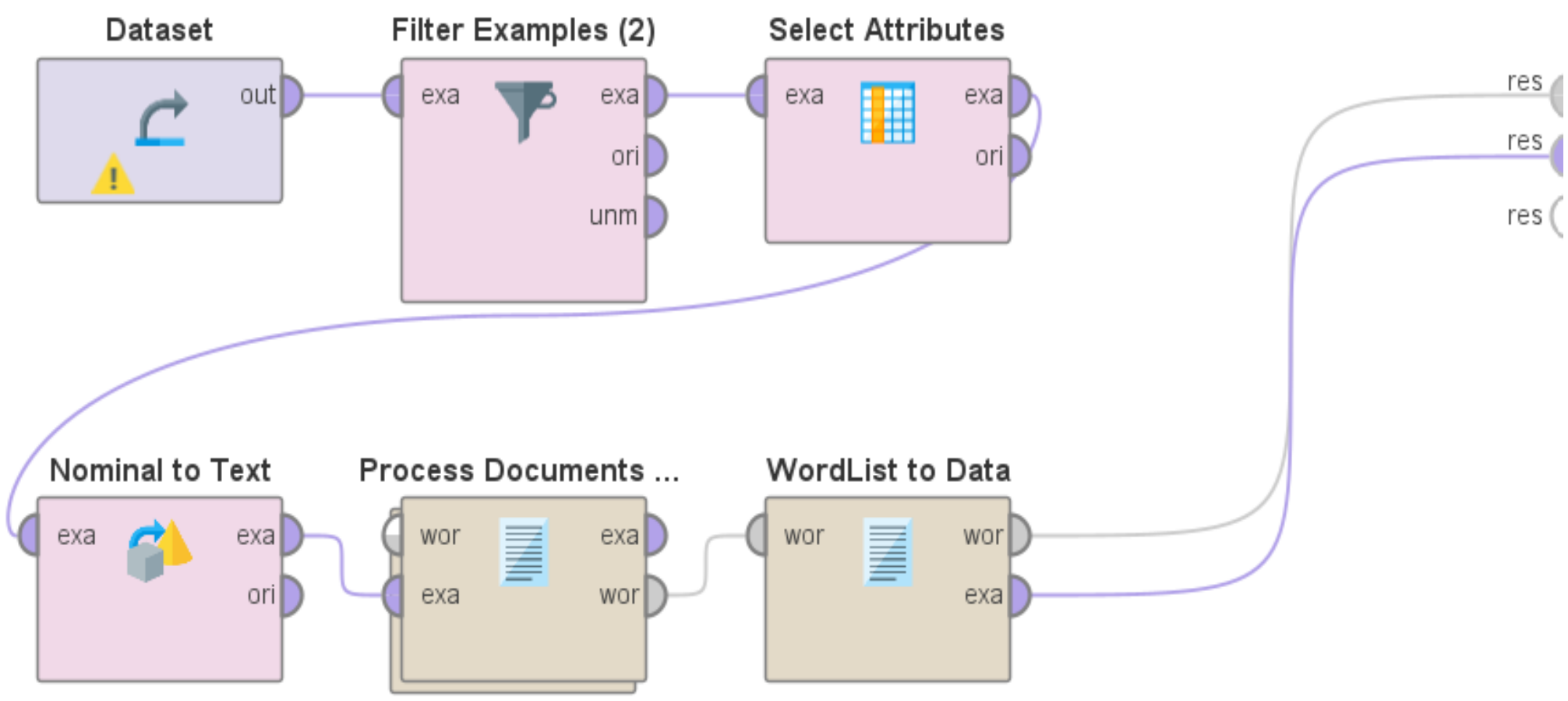

**Figure 3.** RapidMiner process for performing text analysis.

## 4. Results and Discussion

This section is divided into three parts. Section 4.1 presents the results of sentiment analysis of 61,862 tweets that focused on MPox and COVID-19 at the same time. In Section 4.2, the results of the text analysis of the tweets are presented. Specifically, this section reports some of the commonly used hashtags and words that were present in these tweets. Section 4.3 presents a comprehensive comparative study with all the prior works in this field (reviewed in Section 2) to further uphold the scientific contributions of this paper.

### *4.1. Results of Sentiment Analysis*

The sentiment analysis of this set of 61,862 tweets that focused on MPox and COVID-19 at the same time was performed using the VADER approach. The output of the VADER "operator" presented multiple new attributes, and each attribute provided specific information related to the sentiment associated with the tweets that were analyzed. Figure 4 shows the output that was produced by the RapidMiner "process" shown in Figure 2. To avoid presenting an image with 61,862 rows, Figure 4 shows a random selection of 19 rows from the output table. In this Figure, the columns marked in yellow were introduced by the VADER "operator" and were not originally present in the dataset. For each tweet, the VADER approach performed tokenization at first. This is represented in the attributes "Total Tokens" and "Uncovered Tokens" in Figure 4. Thereafter, it captured those tokens from the tweets which expressed either a positive or negative sentiment and then assigned a sentiment score to these respective tokens. This score was assigned on a scale of −4 to +4, where −4 meant highly negative, and +4 meant highly positive. These sets of tokens and their respective sentiment scores comprised the value of the "scoring string" (as shown in Figure 4) for each tweet. Thereafter, for each tweet, the VADER approach grouped all those tokens that had a positive sentiment and computed the sum of the sentiment scores for those tokens. This comprised the "Positivity" value of that tweet. In a similar manner, the VADER approach grouped all those tokens that had a negative sentiment and computed the sum of the sentiment scores for those tokens. This comprised the "Negativity" value of that tweet. The difference between the "Positivity" value and the "Negativity" value was thereafter computed by the VADER approach to display the overall score of the tweet. If this score was negative, the tweet was considered to have an overall negative sentiment. If this score was positive, the tweet was considered to have an overall positive sentiment. Finally, if this score was zero, the tweet was considered to have a neutral sentiment. Based on this analysis, the number of tweets with a positive sentiment was observed to be 29,000, the number of tweets with a negative sentiment was observed to be 19,780, and the number of tweets with a neutral sentiment was observed to be 13,082. This is illustrated in Figure 5.

| Row No. | Score | Scoring String | Negativity | Positivity | Uncovered Tokens | Total Tokens | text |
|---|---|---|---|---|---|---|---|
| 19 | 0.128 | top (0.21) comedy (0.38) lies (-0.46) | 0.462 | 0.590 | 41 | 44 | You know how I know this cou... |
| 20 | 0.179 | growing (0.18) | 0 | 0.179 | 27 | 28 | Monkey pox is the COVID versi... |
| 21 | -0.564 | infected (-0.56) | 0.564 | 0 | 37 | 38 | Bill Gates intentionally infected... |
| 22 | -1.410 | die (-0.74) dumb (-0.59) super (0.74) pissed (-... | 2.154 | 0.744 | 23 | 27 | If I survive leukemia twice and ... |
| 23 | 0.026 | sure (0.33) worried (-0.31) | 0.308 | 0.333 | 62 | 64 | @PaganTri @iva_question @... |
| 24 | -0.282 | want (0.08) super (0.74) strain (-0.05) definite... | 2.282 | 2 | 57 | 67 | If they want me to stay inside &... |
| 25 | 0.359 | like (0.38) increased (0.28) destruction (-0.69)... | 0.692 | 1.051 | 47 | 51 | @debrakidd T cell Depletion d... |
| 26 | -2 | wrong (-0.54) gun (-0.36) violence (-0.79) no ... | 2 | 0 | 17 | 21 | @shawnouten3 And You ainâ... |
| 27 | 0.308 | ok (0.31) | 0 | 0.308 | 25 | 26 | @BoushahriAliaa OK, I quit |
| 28 | -0.026 | yes (0.44) injury (-0.46) | 0.462 | 0.436 | 21 | 23 | Yes it does. Monkey Pox is the... |
| 29 | -0.872 | stop (-0.31) fear (-0.56) | 0.872 | 0 | 23 | 25 | @GBNEWS Covid was a lie a... |
| 30 | -0.846 | no (-0.31) no (-0.31) forget (-0.23) | 0.846 | 0 | 43 | 46 | @amlivemon No equivalency, ... |
| 31 | 0 | | 0 | 0 | 17 | 17 | @Sally718068071 Covid inve... |
| 32 | -2.615 | attack (-0.54) recession (-0.46) recession (-0.4... | 3.128 | 0.513 | 43 | 51 | â˜ï¸□2001 - 9/11 attack |
| 33 | -0.538 | frustration (-0.54) | 0.538 | 0 | 16 | 17 | @DataDrivenMD Covid is tech... |
| 34 | 0.462 | lol (0.46) | 0 | 0.462 | 20 | 21 | Lol. Itâ€™s coming. Part two of ... |
| 35 | 0.821 | please (0.33) thanks (0.49) | 0 | 0.821 | 20 | 22 | Covid numbers are up again, ... |
| 36 | 0.103 | fright (-0.41) welcome (0.51) | 0.410 | 0.513 | 22 | 24 | Covid was really unseen scare... |
| 37 | 0 | | 0 | 0 | 18 | 18 | @MidnightRyderX Monkey pox... |

**Figure 4.** A random selection of 19 rows from the output table generated by RapidMiner.

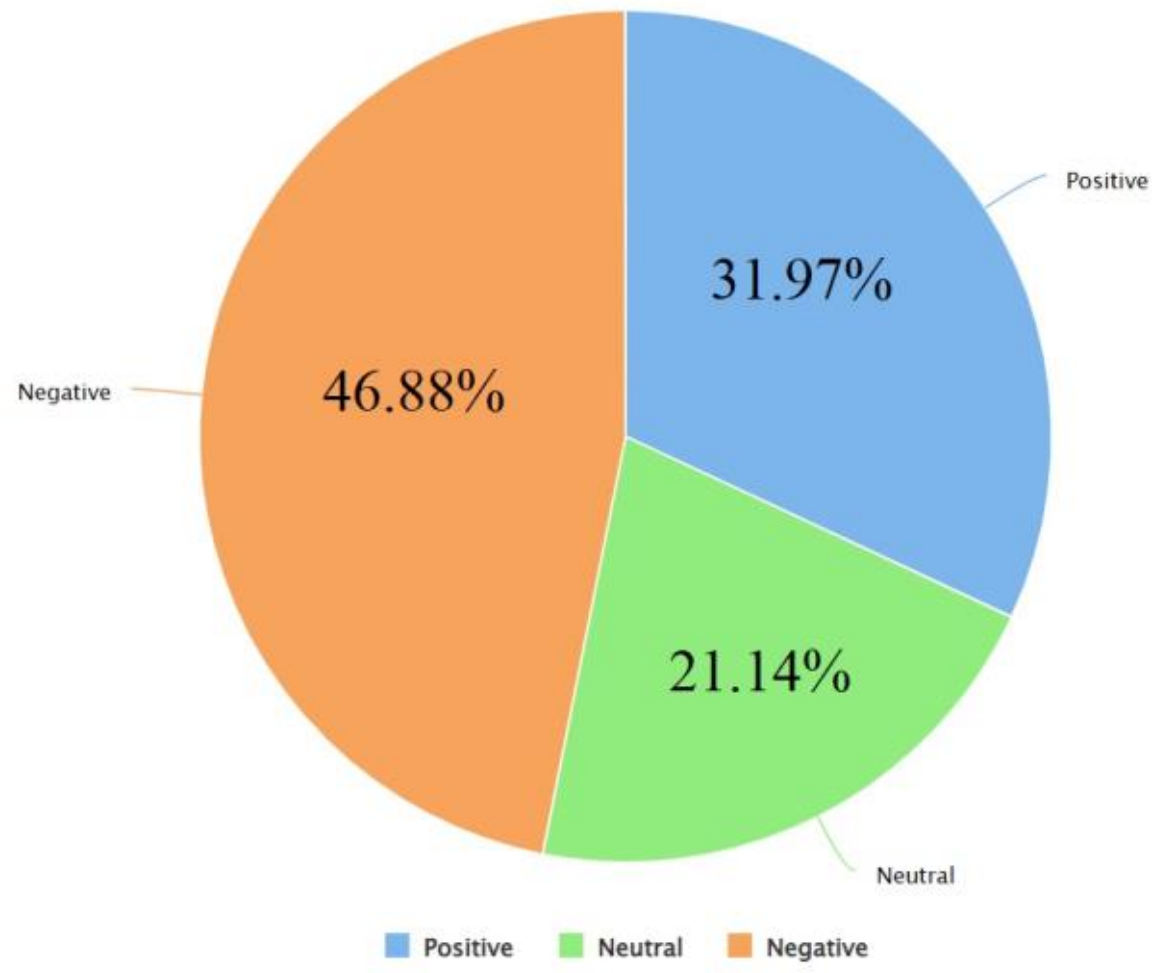

**Figure 5.** Representation of the percentage of positive, negative, and neutral tweets present in this dataset.

According to these findings, it can be concluded that almost half the tweets (46.88%) that focused on COVID-19 and MPox simultaneously had a negative sentiment. It was followed by tweets that had a positive sentiment (31.97%) and tweets that had a neutral sentiment (21.14%), respectively.

### *4.2. Results of Text Analysis*

The results of the text analysis of the tweets are presented in this section. The steps included tokenization, removal of stopwords, and word frequency analysis. Table 2 shows the list of the top 50 hashtags and their frequencies. Here, frequency refers to the number of times each of these hashtags was present in the total number of tweets. The results of tokenization are presented next. In view of the large number of tokens obtained from the set of tweets, this analysis was performed by including the top 100 tokens in terms of their respective frequencies.

**Table 2.** The list of top 50 hashtags and their frequencies in the given tweets.

| Hashtag | Frequency |
|---|---|
| monkeypox | 350 |
| COVID19 | 97 |
| Monkeypox | 88 |
| monkeypox COVID19 | 77 |
| COVID19 monkeypox | 64 |
| COVID | 31 |
| MonkeyPox | 29 |
| SchlongCovid | 27 |
| monkeypoxCOVID | 24 |
| CovidIsNotOver | 21 |
| covidmonkeypox | 21 |
| COVIDmonkeypox | 19 |
| MonkeypoxVirus | 18 |
| monkeypoxCovid_19 | 17 |
| covid19 | 16 |
| COVIDisAirborne | 15 |
| moneypox | 15 |
| monkeypoxcovid | 15 |
| schlongcovid | 15 |
| auspol | 14 |
| COVID19Monkeypox | 13 |
| CovidIsNotOvermonkeypox | 12 |
| MonkeypoxCOVID19 | 12 |
| Covidmonkeypox | 11 |
| Covid19 | 11 |
| LongCovid | 11 |
| covid | 11 |
| covid19monkeypox | 11 |
| Covid_19 | 9 |
| Covid_19monkeypox | 9 |
| LoveIslandUSA | 9 |
| MoneyPox | 9 |
| monkeypoxcovid19 | 9 |
| MonkeypoxCOVID | 8 |
| PrimeMorning | 8 |
| monkeypoxmonkeypox | 8 |
| COVID19ausCOVID19vicWearamask | 7 |
| Covid | 7 |
| Covid19monkeypox | 7 |
| LoveIsland | 7 |
| MedTwitter | 7 |
| MonkeyPoxCOVID19 | 7 |
| monkeypoxCovidIsNotOver | 7 |
| rogerbezanisLetsGoBrandon | 7 |
| CovidMonkeypox | 6 |
| FJB | 6 |
| RussiaUkraine | 6 |
| SmartNews | 6 |
| cdnpoli | 6 |
| covidMonkeypox | 6 |

Table 3 shows these tokens, and a visual representation of the same in the form of a word cloud is shown in Figure 6. As can be seen from this table, several words directly related to these respective viruses are in the list of the top 100 used words. This was expected, as this study focuses on tweets about COVID-19 and MPox. At the same time, the fact that this analysis shows several words that are not directly related to any of these viruses, such as “Polio”, “Biden”, “Ukraine”, “HIV”, “climate”, and “Ebola”, in the list of top 100 most frequent words that featured in these tweets underlines the fact that topics of conversations on Twitter in the context of COVID-19 and MPox also included a high level of interest related to other viruses, President Biden, and Ukraine.

**Table 3.** The list of the top 100 words from these tweets and their respective frequencies.

| Word | Frequency |
|---|---|
| pox | 40,154 |
| monkey | 34,485 |
| Covid | 25,992 |
| covid | 21,385 |
| Monkey | 15,963 |
| COVID | 15,078 |
| Pox | 10,051 |
| monkeypox | 6578 |
| people | 6223 |
| get | 5968 |
| going | 3763 |
| vaccine | 4040 |
| Monkeypox | 3247 |
| got | 3004 |
| time | 2744 |
| know | 2579 |
| shit | 2565 |
| virus | 2540 |
| go | 2331 |
| think | 2286 |
| pandemic | 2226 |
| flu | 2096 |
| want | 2008 |
| polio | 1939 |
| getting | 1985 |
| health | 2005 |
| cases | 2036 |
| spread | 2006 |
| see | 1895 |
| world | 1823 |
| vaccines | 1808 |
| thing | 1614 |
| why | 1586 |
| mask | 1559 |
| years | 1518 |
| make | 1393 |
| disease | 1365 |
| said | 1373 |
| work | 1403 |
| say | 1237 |
| keep | 1167 |
| Polio | 1128 |
| POX | 1133 |
| scared | 1216 |
| fear | 1155 |
| outbreak | 1125 |
| Biden | 1131 |
| Ukraine | 1064 |
| year | 1127 |
| emergency | 1146 |
| stop | 1119 |
| come | 1033 |
| ay | 1092 |
| change | 1017 |
| spreading | 1010 |
| good | 1006 |
| coming | 985 |
| masks | 987 |
| global | 973 |
| bad | 954 |
| HIV | 943 |
| climate | 925 |
| trying | 897 |
| Why | 940 |
| day | 898 |
| MONKEY | 862 |
| news | 903 |
| vaccinated | 893 |
| cause | 862 |
| stay | 827 |
| vax | 1001 |
| government | 820 |
| care | 844 |
| safe | 810 |
| else | 769 |
| CDC | 822 |
| made | 785 |
| days | 802 |
| country | 765 |
| shot | 979 |
| Flu | 755 |
| sick | 765 |
| believe | 750 |
| case | 758 |
| risk | 791 |
| start | 717 |
| corona | 727 |
| catch | 736 |
| control | 753 |
| thought | 711 |
| saying | 725 |
| look | 706 |
| diseases | 720 |
| Ebola | 714 |
| moneypox | 689 |
| kids | 744 |
| life | 699 |
| sex | 756 |
| give | 695 |
| Lol | 691 |

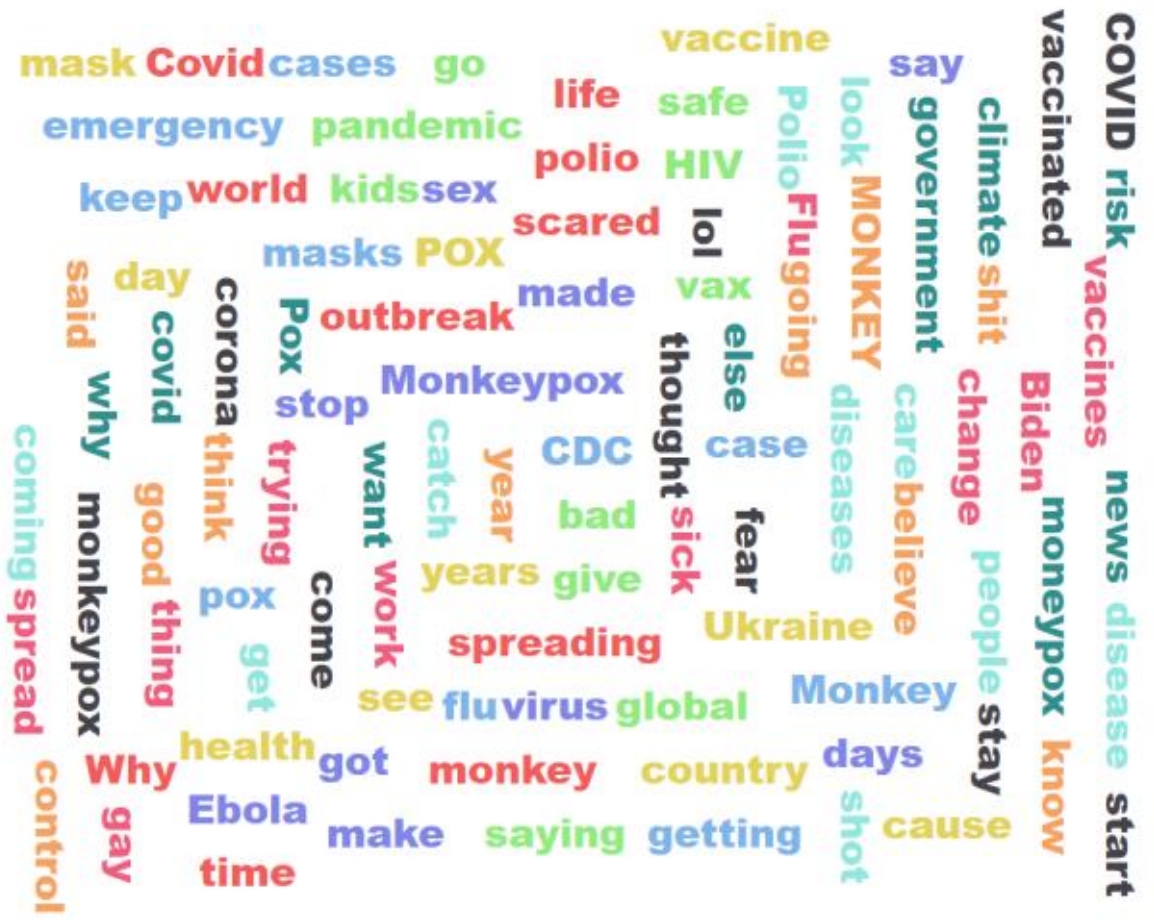

**Figure 6.** Representation of some of the most frequently used words in these tweets in the form of a word cloud.

## 4.3. Comparative Study with Prior Works

This section presents a comparative study with prior works in this field (reviewed in Section 2). This comparative study is represented in Table 4. As can be seen from Table 4, the work presented in this paper is the first work in this area that focuses on sentiment analysis of tweets that focused on COVID-19 and MPox at the same time.

**Table 4.** Comparative study with prior works in this field.

| Work | Sentiment Analysis of Tweets about COVID-19 | Sentiment Analysis of Tweets about MPox |
|---|---|---|
| Vijay et al. [77] | ✓ | |
| Mansoor et al. [78] | ✓ | |
| Pokharel [79] | ✓ | |
| Chakraborty et al. [80] | ✓ | |
| Shofiya et al. [81] | ✓ | |
| Basiri et al. [82] | ✓ | |
| Cheeti et al. [83] | ✓ | |
| Ridhwan et al. [84] | ✓ | |
| Tripathi [85] | ✓ | |
| Situala et al. [86] | ✓ | |
| Gupta et al. [87] | ✓ | |
| Alanezi et al. [88] | ✓ | |
| Dubey [89] | ✓ | |
| Rahman et al. [90] | ✓ | |
| Ainlet et al. [91] | ✓ | |
| Slobodin et al. [92] | ✓ | |
| Zou et al. [93] | ✓ | |
| Alhuzali et al. [94] | ✓ | |
| Hussain et al. [95] | ✓ | |
| Liu et al. [96] | ✓ | |
| Hu et al. [97] | ✓ | |
| Khan et al. [98] | ✓ | |
| Ahmed et al. [99] | ✓ | |
| Lin et al. [100] | ✓ | |
| Jang et al. [101] | ✓ | |
| Tsao et al. [102] | ✓ | |
| Griffith et al. [103] | ✓ | |
| Chum et al. [104] | ✓ | |
| Kothari et al. [105] | ✓ | |
| Barkur et al. [106] | ✓ | |
| Afroz et al. [107] | ✓ | |
| Hota et al. [108] | ✓ | |
| Venigalla et al. [109] | ✓ | |
| Paliwal et al. [110] | ✓ | |
| Zhou et al. [111] | ✓ | |
| Lamsal et al. [112] | ✓ | |
| Zhou et al. [113] | ✓ | |
| de Melo et al. [114] | ✓ | |
| Brum et al. [115] | ✓ | |
| de Sousa et al. [116] | ✓ | |
| Iparraguirre-Villanueva et al. [117] | | ✓ |
| Mohbey et al. [118] | | ✓ |
| Farahat et al. [119] | | ✓ |
| Sv et al. [120] | | ✓ |
| Bengesi et al. [121] | | ✓ |
| Dsouza et al. [122] | | ✓ |
| Zuhanda et al. [123] | | ✓ |
| Cooper et al. [124] | | ✓ |
| Ng et al. [125] | | ✓ |
| Thakur [this work] | ✓ | ✓ |

## 5. Conclusions

The big data of Twitter conversations holds the potential for the inference of the views, opinions, perspectives, mindsets, sentiments, and feedback of the general public towards pandemics, epidemics, viruses, and diseases. This has attracted the attention of researchers in the fields of computer science, big data, data science, epidemiology, healthcare, medicine, and their interrelated areas in the last few years. Various forms of analysis of this big data, such as sentiment analysis, hashtag analysis, and frequent-keyword analysis, can be seen in prior works in this field that focused on studying tweets involving some of the virus outbreaks of the past, such as Ebola, E-Coli, Dengue, Human Papillomavirus, Middle East Respiratory Syndrome, Measles, Zika virus, H1N1, influenza-like illness, swine flu, flu, Cholera, COVID, Listeriosis, cancer, Liver Disease, Inflammatory Bowel Disease, kidney disease, lupus, Parkinson's, Diphtheria, and West Nile virus. The recent outbreaks of COVID-19 and MPox have escalated the use of Twitter for conversations related to these respective viruses. While there have been a few works published in the last few months that focused on performing sentiment analysis of tweets related to either COVID-19 or MPox, none of the prior works in this field thus far focused on the analysis of tweets focusing on both COVID-19 and MPox at the same. To address this research gap, this study presents the findings from a comprehensive sentiment analysis of 61,862 tweets that focused on MPox and COVID-19 at the same time. The VADER approach was used to perform the sentiment analysis. The results show that almost half the tweets (46.88%) involving COVID-19 and MPox had a negative sentiment. It was followed by tweets that had a positive sentiment (31.97%) and tweets that had a neutral sentiment (21.14%), respectively. This study also presents the findings from hashtag analysis and keyword analysis of these tweets. The top 50 hashtags that featured in all these tweets were detected and are presented in this paper. The top 100 most frequently used words that featured in all these tweets were also detected using concepts of tokenization. The findings of frequent word analysis show that some of the commonly used words directly referred to either or both of these viruses. In addition to this, the presence of words such as "Polio", "Biden", "Ukraine", "HIV", "climate", and "Ebola" in the list of the top 100 most frequent words indicate that topics of conversations on Twitter in the context of COVID-19 and MPox also included a high level of interest related to other viruses, President Biden, and Ukraine. A limitation of this study is that the data preprocessing and analysis did not involve the detection and elimination of tweets posted by bot accounts on Twitter. Future work would involve addressing this limitation and collecting more tweets over the next months to repeat this study, with an aim to infer and analyze any potential evolution or trends of public sentiment related to these viruses over the course of time.

**Author Contributions:** Conceptualization, N.T.; methodology, N.T.; software, N.T.; validation, N.T.; formal analysis, N.T.; investigation, N.T.; resources, N.T.; data curation, N.T.; writing—original draft preparation, N.T.; writing—review and editing, N.T.; visualization, N.T.; project administration, N.T.; funding acquisition, not applicable.

**Funding:** This research received no external funding.

**Institutional Review Board Statement:** Not applicable. Ethical review and approval were waived for this study due to the following reason: the manuscript meets ethical standards and regulations since this research meets one of the IRB 45 CFR 46.101(b) requirements in Category 4—Research involving the collection or study of existing data, documents, records, pathological specimens, or diagnostic specimens, if these sources are publicly accessible or if the information is recorded by the investigator in such a way that subjects cannot be directly or indirectly identified.

**Data Availability Statement:** The data analyzed in this study are publicly available at https://dx.doi.org/10.21227/16ca-c879 (accessed on 27 March 2023).

**Conflicts of Interest:** The author declares no conflict of interest.